\documentclass[preprintnumbers,nofootinbib,noshowpacs,eqsecnum,prd,superscriptaddress,letterpaper]{revtex4-2}
\pdfoutput=1
\usepackage{graphicx}
\usepackage{amsmath,amssymb}
\usepackage{url}
\usepackage{multirow}
\usepackage{hyperref,color}
\usepackage{slashed}
\usepackage{rotating}
\usepackage{afterpage}
\usepackage{xcolor}
\usepackage{ifthen}
\usepackage{mciteplus}
\usepackage{multirow}
\usepackage{color}
\usepackage{pifont}
\usepackage{float}

\usepackage{caption}
\usepackage{soul}

\usepackage{subcaption}
\usepackage{hyperref}
\hypersetup{
	colorlinks=true,       
	linkcolor=blue,        
	citecolor=red,        
	urlcolor=blue         
}
\usepackage[capitalize]{cleveref}

\begin{document}
\renewcommand{\baselinestretch}{1.5}
\preprint{YITP-SB-2026-07}

\title{Exploring neutrino loss with difusse astrophysical neutrino fluxes}

\author{Ivan Esteban}
\email{ivan.esteban@ehu.eus}
\affiliation{Department of Physics, University of the Basque Country
  UPV/EHU, PO Box 644, 48080 Bilbao, Spain}
\affiliation{EHU Quantum Center, University of the Basque Country
  UPV/EHU}

\author{Alberto M.~Gago}
\email{agago@pucp.edu.pe}
\affiliation{Sección Física, Departamento de Ciencias, Pontificia
  Universidad Católica del Perú, Apartado 1761, Lima, Perú}

\author{M.~C.~Gonzalez-Garcia}
\email{concha.gonzalez-garcia@stonybrook.edu}
\affiliation{C.N. Yang Institute for Theoretical Physics, Stony Brook University, Stony Brook NY11794-3849,  USA}
\affiliation{Departament  de  Fisica  Quantica  i  Astrofisica
 and  Institut  de  Ciencies  del  Cosmos,  Universitat
 de Barcelona, Diagonal 647, E-08028 Barcelona, Spain}
\affiliation{Instituci\'o Catalana de Recerca i Estudis Avancats (ICREA)
Pg. Lluis  Companys  23,  08010 Barcelona, Spain.}
\author{Gabriel D.~Zapata}
\email{gabriel.zapata@pucp.edu.pe}
\affiliation{Sección Física, Departamento de Ciencias, Pontificia
  Universidad Católica del Perú, Apartado 1761, Lima, Perú}

\date{\today}

\begin{abstract}
We study the sensitivity of the diffuse high-energy neutrino flux
observed in IceCube to new-physics effects resulting in an exponential
flux attenuation along the trajectory, such as invisible neutrino
decay or new interactions with the background encountered during
propagation. We argue that, even though the sources and production
redshifts of these astrophysical neutrinos are unknown, conservative
energy-conservation arguments allow to severely constrain neutrino
loss in most scenarios beyond the strongest existing bounds. By
performing a fit to the High-Energy Starting Events from IceCube, we
quantify the bounds and study their variation with the energy
dependence of the attenuation, the assumed redshift distribution of
the neutrino sources, and whether the attenuation affects neutrinos
exclusively or no.  We also show that including an energy-dependent
attenuation at the level allowed in the fit may impact the
determination of the spectral index of the diffuse flux.
\end{abstract}

\maketitle

\section{Introduction}
Ultra-high-energy cosmic rays (UHECRs) are most likely produced and
accelerated in extreme astrophysical environments (see, {\it e.g.},
Refs.~\cite{Kotera:2011cp, Anchordoqui:2018qom} for a review).  As
such, they are expected to interact with the radiation and matter
surrounding the sources, as well as with background radiation in their
way to Earth. The outcome of such processes is a flux of high-energy
(HE) neutrinos detectable at Earth.

In the Standard Model of Particle Physics (SM) neutrinos interact very
weakly with matter and radiation, and the HE neutrino flux is expected
to arrive at Earth isotropically and with negligible
attenuation. Conversely, this rather robust SM prediction implies that
the observation of HE neutrinos has potential sensitivity to
new-physics effects affecting their propagation, in particular those
that would attenuate the neutrino flux arriving at Earth.  This
includes, for example, the decay of neutrinos totally or partly into
non-interacting states~\cite{Valera:2024buc, Bustamante:2016ciw,
  Shoemaker:2015qul, Denton:2018aml}; or interactions with the
environment, such as Dark Matter (DM) particles
~\cite{Ferrer:2022kei,
  Cline:2022qld, Cline:2023tkp, Bertolez-Martinez:2025trs,
  Mondol:2025uuw, Zapata:2025huq, Esteban:2025wbv}
or an stochastic
environment sourced, for example, by the quantum foam in a quantum
theory of gravity. This last case generically results in a loss of
coherence~\cite{Hawking:1976ra, Ellis:1983jz, Giddings:1988cx,
  Addazi:2021xuf}, for example via interactions of neutrinos with
virtual black holes, and in some scenarios it can lead to neutrino
loss~\cite{Stuttard:2020qfv}.

In general, all these effects result in the exponential attenuation of
the neutrino flux arriving at Earth with a (generically
energy-dependent) exponent that grows with the distance traveled by
the neutrino. It is this distance dependence that has made the known
astrophysical neutrino sources the natural testbeds for these
effects. This reduces the choices to the observed MeV neutrino fluxes
from the Sun~\cite{Cleveland:1998nv, Kaether:2010ag,
  Abdurashitov:2009tn, Hosaka:2005um, Cravens:2008aa, Abe:2010hy,
  Super-Kamiokande:2023jbt, Aharmim:2011vm, BOREXINO:2018ohr} and
SN1987A~\cite{Kamiokande-II:1987idp,Alekseev:1988gp,Bionta:1987qt},
and the two identified HE neutrino sources NGC
1068~\cite{IceCube:2022der} and TXS
0506+056~\cite{IceCube:2018cha}. At the same time, this makes the
studies intrinsically source-model dependent.  In that respect, solar
neutrino models are the best understood and tested. Thus, despite
their lower energies and closer distance, solar neutrinos have been
used to place bounds on decay lifetimes~\cite{Beacom:2002cb,
  Berryman:2014qha, Funcke:2019grs} as well as on the effect of
quantum decoherence on the oscillation pattern (though not in the
neutrino-loss scenario). SN1987A data has been exploited to constraint
neutrino decay~\cite{Ivanez-Ballesteros:2023lqa,
  Martinez-Mirave:2024hfd}, and most recently neutrino quantum
decoherence in the neutrino-loss scenario~\cite{Ternes:2025mys}. As
mentioned above, HE neutrinos have been employed to test partial
attenuation due to decay~\cite{Valera:2024buc, Bustamante:2016ciw,
  Shoemaker:2015qul, Bustamante:2016ciw, Denton:2018aml} and loss due
to DM interactions~\cite{Ferrer:2022kei, Cline:2022qld, Cline:2023tkp,
  Bertolez-Martinez:2025trs, Mondol:2025uuw, Zapata:2025huq,
  Esteban:2025wbv}.

In this work, instead of fluxes from specific sources, we explore the
implications of generic energy- and distance-dependent neutrino loss
effects on a diffuse HE flux such as the one observed by the IceCube
neutrino telescope~\cite{IceCube:2013low, IceCube:2020wum,
  IceCube:2023sov}. Our starting point is the observation that, even
though the specifics of the sources and production redshifts of these
astrophysical neutrinos are unknown, conservative energy-conservation
arguments allow to severely constrain exponentially growing neutrino
losses in most scenarios, beyond bounds derived from specific
sources. To this end, we introduce in Sec.~\ref{sec:form} our
formalism to compute the high-energy astrophysical neutrino flux
including neutrino loss (Sec.~\ref{sec:form1}), and we argue how the
observed diffuse flux allows to constraint neutrino loss in
Sec.~\ref{sec:form2}, with the assumptions and analysis choices
described Sec.~\ref{sec:analysis_choices}. In Sec.~\ref{sec:predic},
we show the dependence of the predicted diffuse fluxes on the
different assumptions entering in the
analysis. Section~\ref{sec:results} contains the results of our
detailed analysis of the High-Energy Starting Events (HESE) from
IceCube~\cite{IceCube:2020wum} (see also information
in Appendix ~\ref{sec:fit-params}), in the form of constraints on the
neutrino loss exponent. We quantify the variation of the bounds with
the energy dependence of attenuation, the source evolution, and the
universality of the effect; and we show the impact of attenuation in
determining the spectral index of the diffuse flux.  We summarize our
conclusions in Sec.~\ref{sec:conclu}.

\section{Formalism}
\label{sec:form}
In this Section, we describe our formalism to compute the high-energy
astrophysical neutrino flux including neutrino loss. We also argue how
the observed diffuse flux allows to constraint neutrino loss, despite
the sources of this flux being unknown. In short, energy-conservation
arguments, combined with the large distances to astrophysical objects,
set conservative bounds on the amount of high-energy neutrinos and
their traveled distance. Exponentially large neutrino loss would be in
conflict with the observation of a diffuse neutrino flux at IceCube.

\subsection{Evolution of the neutrino flux}
\label{sec:form1}
In order to compute the astrophysical neutrino flux arriving at Earth
including generic energy-dependent neutrino loss, one must take into
account that the production redshift of these neutrinos is unknown and
potentially large. Therefore, the expansion of the Universe must be
taken into account.

The evolution of the comoving density of neutrinos of mass eigenstate
$i$ per unit energy $E$ is governed by the following transport
equation~\cite{Berezinsky:2005fa}
\begin{equation}
    \frac{\partial n_i(E, t)}{\partial t} = \mathcal{I}_i(E, t) +
    \frac{\partial}{\partial E} \left[ H(t) E n_i(t, E) \right] - \Gamma_i (E) n_i(E, t) \, ,
\end{equation}
with $t$ cosmological time, $\mathcal{I}_i$ the comoving rate of
neutrino production with flavour $i$, $H$ the Hubble rate, and
$\Gamma_i$ the neutrino loss rate. Physically, the first term captures
neutrino production, the second term neutrino energy redshift due to
the expansion of the Universe, and the third term the neutrino loss
that is responsible for the attenuation of the neutrino flux.
Throughout this work, when we mention neutrinos of a given type we
always refer to the sum of neutrinos and antineutrinos of that
type. We assume that the relevant propagation scales are much larger
than neutrino oscillation lengths, which averages out flavour
oscillations and allows to directly follow the evolution of mass
eigenstates (on top of that, mass eigenstates are expected to decohere
over astrophysical scales~\cite{Nussinov:1976uw}).

By writing it in terms of cosmological redshift $z$, the transport
equation can be solved to obtain the neutrino flux at Earth per unit
energy, area, time, and solid angle; $\phi_i \equiv
\frac{\mathrm{d}N_i}{\mathrm{d}E\mathrm{d}A\mathrm{d}t\mathrm{d}\Omega}
= \frac{1}{4\pi} n_i(E, z=0)$ as~\cite{Ahlers:2009rf}
\begin{equation}
  \begin{split}
\phi_i & = \int_{z_\mathrm{min}}^\infty \frac{\mathrm{d}z}{4\pi H(z)}
\mathcal{I}_i((1+z) E, z) e^{-\tau(E, z)} \, ,
\label{eq:nu_flux_evolved}
  \end{split}
\end{equation}
where $z_\mathrm{min}$ is the smallest cosmological redshift at which
neutrinos that can be detected at Earth are produced (more on this
below), and we have introduced an effective optical depth
\begin{equation}
  \tau(E, z) \equiv \int_0^z \frac{\mathrm{d}z'}{H(z')(1+z')}
  \Gamma_i((1+z') E) \, ,
  \label{eq:tau_eff}
\end{equation}
that can be understood as the integral of the neutrino loss rate
$\Gamma$ over the lookback time $\mathrm{d}t_\mathrm{L} =
\frac{\mathrm{d}z'}{H(z')(1+z')}$, from the emission redshift $z$ to
the detection redshift $z=0$. In all cases, the prefactors $(1+z)$
multiplying neutrino energy are due to cosmological redshift. The flux
of the mass eigenstate $i$, $\phi_i$, can be immediately converted to
the flux of flavour $\alpha$, $\phi_\alpha$, as
\begin{equation}
  \phi_\alpha \equiv \frac{\mathrm{d}N_\alpha}{\mathrm{d}E\mathrm{d}A\mathrm{d}t\mathrm{d}\Omega} = \sum_i |U_{\alpha i}|^2 \phi_i\, ,
\end{equation}
with $U$ the leptonic mixing matrix.

The diffuse astrophysical neutrino flux is expected to originate at
many unresolved sources with comoving density
$\rho_\mathrm{src}(z)$. We thus parametrize the neutrino production
rate as
\begin{equation}
  \mathcal{I}_i(E, z) = \rho_\mathrm{src}(z) 
  \sum_\beta |U_{\beta i}|^2 \left.\frac{\mathrm{d} N_\beta}{\mathrm{d}E
    \mathrm{d}t}\right|_\mathrm{S}(E)\, ,
  \label{eq:nu_prod_rate}
\end{equation}
where $\left.\frac{\mathrm{d} N_\beta}{\mathrm{d}E \mathrm{d}t}\right|_\mathrm{S}$ is the
amount of neutrinos with flavour $\beta$ that are produced per unit
energy and unit time in each source. In \cref{sec:analysis_choices}
below, we detail our choices for these functions.

For the sake of concreteness, we assume that neutrino loss 
depends on energy as a power law
\begin{equation}
\Gamma_i(E) = \gamma_{0,i} \left(\frac{E}{E_0}\right)^n \, ,
\label{eq:Gam}
\end{equation}
where $E_0$ is a reference energy. For easy comparison with existing
literature~\cite{Stuttard:2020qfv, Ternes:2025mys}, we set
$E_0=1$\,GeV and consider the benchmark values $n=-2,-1,0,1,2$.

This generic parametrization has been used in the literature (see
Refs.~\cite{Stuttard:2020qfv, Ternes:2025mys} and references therein)
when describing neutrino loss in the context of quantum gravity, with
$n=2$ being of particular interest as this value has been predicted in
string-theory models~\cite{Ellis:1996bz,Ellis:1997jw}. The same
parametrization also describes the decay of neutrinos with mass $m_i$ into
invisible (this is, non-detectable) states with lifetime $\tau_i$.
In this case
\begin{equation}
  \gamma^{\rm inv-dec}_{0,i}=\frac{m_i}{\tau_i E_0} \;,\hspace*{1cm}
  n^{\rm inv-dec}=-1 \;,
\label{eq:pardec}
\end{equation}
where the energy dependence results from time dilation.  The
parametrization can also be applied to neutrino interactions with a
background of relic particles.  For example, for interactions with
Dark Matter particles with mass $M_{\rm DM}$ and present-day average
density $\rho^{\rm DM}_0$, if the neutrino-Dark Matter interaction
cross section depends on energy as $\sigma^{\rm
  \nu-DM}(E)=\sigma_{0,i} (E/E_0)^{n_{\sigma}}$ in the relevant
neutrino energy range~\footnote{In general, $n_\sigma$ depends on the
masses and spins of Dark Matter and the interaction mediator, see
Ref.~\cite{Arguelles:2017atb} for explicit examples.}
\begin{equation}
 \gamma^{\rm \nu-DM}_{i,0}=\frac{\rho_0\, \sigma_{0,i}}{M_{\rm DM}} \;,\hspace*{1cm}
 n^{\rm \nu-int}= n_{\sigma} \;,
\label{eq:pardm} 
\end{equation}
or equivalently $\gamma^{\rm \nu-bck}_{i,0}=\sigma_{0,i} n_0^{\rm bck}$
for interactions with any relic background with average
current number density $n_0^{\rm bck}$ and power-law interaction cross section.
In principle, these scenarios introduce additional redshift dependence
 in Eq.~\eqref{eq:tau_eff}, as the background Dark-Matter density and
 hence the neutrino loss rate scale as $(1+z')^3$. This would
 effectively modify the $n$-dependence of the $(1+z')^n$ factor
 stemming from neutrino energy redshift in
 Eq.~\eqref{eq:tau_eff}. However, as we show in
 Eq.~\eqref{eq:tau_approx} and Fig.~\ref{fig:rdec} below,
once the $E^{n_{\sigma}}$ depence is factorized out, the impact
 of the residual $n$-dependence on the overall attenuation is
 very weak. Thus, the bounds derived by the simple recipe in
 \cref{eq:pardm} hold up to a factor of at most ${\cal O}(3)$. An
 additional possible effect of this scenario that is not captured in
 our framework is that, strictly speaking, if the scattering process
 is elastic there is no \emph{neutrino loss} but \emph{neutrino energy
 degradation}. The underlying assumption for our framework to apply is
 that the final-state neutrino does not have enough energy to
 contribute to the HESE events.
 
Finally, for simplicity we work under the assumption that neutrino
loss is flavour-universal, so that we can drop the $i$ dependence
of $\gamma_0$. While this assumption is well-justified for
quantum-gravity effects or interactions with Dark Matter or other
relic backgrounds, it is less so for neutrino decay, where in most
scenarios $m_i/\tau_i$ is expected to depend on $i$. In these cases,
the bounds we derive assuming flavour universality can always be
conservatively applied to the state with smallest $\gamma_{0,i}$.  
\subsection{Constraints on the amount of accumulated neutrino loss}
\label{sec:form2}
As mentioned above, the sources and production redshifts of
astrophysical neutrinos are unknown. In principle, this should hinder
setting limits on neutrino loss with diffuse astrophysical neutrino
observations.  However, here we argue that conservative
energy-conservation arguments allow to constrain exponentially large
neutrino loss such as the one we consider in this work. We will derive
these constraints for two limiting scenarios, which we will label as
\emph{Only $\nu$ Attenuation} and \emph{All Particle Attenuation}.

In the first scenario, we make use of the fact that high-energy
astrophysical neutrinos are expected to be produced when high-energy
protons interact with ambient light, producing pions that decay to
neutrinos. Assuming that protons can escape the sources, and given the
observed high-energy cosmic-ray flux, Ref.~\cite{Waxman:1998yy} placed
a conservative upper bound on the neutrino flux that should reach
Earth
\begin{equation}
  \sum_\alpha E^2 \phi_\alpha \lesssim 6 \times
  10^{-8}\,\mathrm{GeV\,cm^{-2}\,s^{-1}\,sr^{-1}}\, ,
\label{eq:lim1}    
\end{equation}
also known as the Waxman-Bahcall (WB) bound. The actual flux is probably
smaller, as the WB bound assumes, among others, strong redshift evolution and a
proton-dominated cosmic-ray composition~\cite{Bahcall:1999yr}. 

Following Ref.~\cite{Esteban:2025wbv}, we argue that one can place a
limit on neutrino loss imposing two requirements. First, that the
neutrino flux that would reach Earth without loss does not violate the
WB bound, which limits the neutrino production rate. Second, that the
neutrino flux at Earth including loss matches the observations at
IceCube, which, together with the previous limit, bounds the neutrino
loss rate. Even if the WB bound is understood as an order-of-magnitude
estimation more than an accurate limit (for example, it can be relaxed
if a large fraction of neutrino sources are optically thick to
protons), neutrino loss is exponential. Hence, neutrino loss stronger
than our limit would require, to be consistent with observations,
unattenuated fluxes that exponentially violate the WB bound.

Since the WB bound depends on energy, we impose it for all neutrino
energies in our analysis ($60\,\mathrm{TeV} \lesssim E \lesssim
10^3\,\mathrm{TeV}$, see below).  As mentioned
above, in what follows we refer to this scenario as \emph{Only $\nu$
Attenuation}.

In principle, such limit could be circumvented if the observed
cosmic-ray flux that drives the WB bound is also subject to strong
attenuation, either at the sources or during propagation due to
mechanisms similar to neutrino loss. This is what we define as the
\emph{All Particle Attenuation} scenario. In this case, one can still
place a conservative limit by requiring the energy density contained
in the diffuse neutrino flux before loss to be smaller than the total
energy density contained in all galaxies in the Universe.

Following Ref.~\cite{Fukugita:2004ee}, we consider the energy density
in galaxies to be 1\% of the critical density of the Universe $\rho_c
\simeq 5 \times 10^{-6}\,\mathrm{GeV\,cm^{-3}}$. This sets a limit on
the present-day diffuse high-energy neutrino energy density
\begin{equation}
   \sum_\alpha E^2 n_\alpha(E, z=0) \lesssim 5 \times
   10^{-8}\,\mathrm{GeV\,cm^{-3}} \, ,
\end{equation}
or, equivalently, 
\begin{equation}
  \sum_\alpha E^2 \phi_\alpha \lesssim 1.2 \times 10^2\,\mathrm{GeV\,cm^{-2}\,s^{-1}\,sr^{-1}} \, ,
\label{eq:lim2}  
\end{equation}
which is about 9 orders of magnitude weaker than the WB bound. While
this bound is \emph{extremely} conservative (to saturate it, neutrino
production processes should convert \emph{all} the galactic rest mass
into neutrinos), neutrino loss is exponential, so the derived limits
do not differ that much from those obtained employing the WB
bound. Since this bound depends on energy, we impose it for all
neutrino energies in our analysis ($60\,\mathrm{TeV} \lesssim
E\lesssim 10^3\,\mathrm{TeV}$, see below). As mentioned above, in what
follows we refer to this scenario as \emph{All Particle Attenuation}.

In both scenarios, we conservatively assume that the energy in
high-energy neutrinos is completely lost. If this energy gets
transferred to other observable particles such as photons or
lower-energy neutrinos, the observation of these would impose
additional constraints.

\subsection{Analysis choices}
\label{sec:analysis_choices}
Here, we provide details on our choices for the different quantities
entering the neutrino production rate, \cref{eq:nu_prod_rate}; and
propagation over cosmological scales,
\cref{eq:nu_flux_evolved,eq:tau_eff}.
\begin{itemize}
\item  \emph{Redshift distribution of high-energy diffuse astrophysical
neutrino sources}: $\rho_\mathrm{src}(z)$. We generically write
\begin{equation}
  \rho_\mathrm{src}(z)=\rho_0\, F(z)\, ,
\end{equation}
with $F(z)$ a dimensionless function satisfying $F(0)=1$. As $F(z)$ is
unknown, we consider two scenarios
\begin{enumerate}
  \item $\rho_\mathrm{src}$ follows the star formation
    rate~\cite{Hopkins:2006bw, Yuksel:2008cu},
    which places most
    sources around $z = 1$
    (more on this later). Explicitly, we use
    the parametrization in Ref.~\cite{Yuksel:2008cu}
    \begin{equation}
      \begin{split}
      F(z)  \equiv F_\mathrm{SFR}(z)  = \biggl[ (1 + z)^{{a}{\eta} } + \left(\frac{1 + z}{B}\right)^{{b}{\eta}} +\left(\frac{1 + z}{C}\right)^{{c}{\eta} } \biggr]^{1/\eta} \, ,
      \end{split}
      \label{eq:sfr}
    \end{equation}
    where $a = 3.4$, $b = -0.3$, $c = -3.5$, $\eta = -10$, $B =
    2^{1-a/b} \simeq 5000$, and $C =2^{(b-a)/c} \, 5^{1-b/c}\simeq 9$.
  \item $\rho_\mathrm{src}$
   follows the distribution of BL-Lac
    objects, a type of AGN. This falls rapidly with redshift, which
    places more sources closer to Earth. We extract the distribution from Fig.~5
    of Ref.~\cite{Capanema:2020oet} (in turn obtained from
    Ref.~\cite{Ajello:2013lka}), approximating it as
    \begin{equation}
      F(z) \equiv F_\mathrm{BL-Lac}(z)=(1+0.7 z^3+ 0.1 z^5)^2
      e^{-z/0.19} \, .
      \label{eq:bl-lac}
    \end{equation}
\end{enumerate}
$\rho_0$ contributes to the overall flux normalization. In our
analysis below, we allow it to float subject to the constraint in
\cref{eq:lim1} and \eqref{eq:lim2} for the \emph{Only $\nu$
Attenuation} and the \emph{All Particle Attenuation}
scenarios, respectively (see \cref{eq:norm} below).
  
\item \emph{Smallest redshift of neutrino sources}: $z_\mathrm{min}$
  in \cref{eq:nu_flux_evolved}. Generically, one would expect that it
  can be safely set to zero because the contribution of extremely low
  redshifts to the diffuse flux is negligible once the entire history
  of the Universe is integrated over. In other words, the amount of
  neutrinos that have traveled small distances before reaching the
  Earth originate from a relatively small volume. Once the volume of
  the whole observable Universe is taken into account, their
  contribution to the diffuse flux is negligible.  However, this stops
  being the case in the \emph{All Particle Attenuation} scenario,
  where the high-energy astrophysical neutrino flux comprises a
  considerable fraction of the total energy density of the
  Universe. In this case, we set a conservative bound using the
  distance to the closest galaxy, M31, $z_\mathrm{min}=2\times 10^{-4}$. We
  also study the dependence of our results on this choice.

\item \emph{Neutrino production spectrum}: we assume a power law (as
favored by data~\cite{IceCube:2020wum})
\begin{equation}
  \left.\frac{\mathrm{d}N_\beta}{\mathrm{d}E \mathrm{d}t}\right|_\mathrm{S}(E)=
f_\beta \Phi_{0}
\left(\frac{E}{1\,\mathrm{TeV}}\right)^{-\gamma_\mathrm{astro}} \, ,
\label{eq:PL}
\end{equation}
with $\gamma_\mathrm{astro}$ the spectral index, $\Phi_0$ an overall
normalization factor, and $f_{\beta S} \in [0, 1]$ the fractional
flavour composition at the source.

$\gamma_\mathrm{astro}$ is one of the parameters to be fitted in the
analysis. The normalization factor, $\Phi_{0}$, is also a free
parameter to be fitted in the analysis subject to the constraint in
\cref{eq:lim1} and \eqref{eq:lim2} for the \emph{Only $\nu$
Attenuation} and the \emph{All Particle Attenuation}
scenarios, respectively (see \cref{eq:norm} below). Finally, we consider two scenarios for
$f_{\beta S}$. In the standard $\pi$-decay dominance production,
  \begin{equation}
    \left(f_{eS},f_{\mu S},f_{\tau S}\right)= \left( \frac{1}{3}, \frac{2}{3}, 0 \right)  \;.
    \label{eq:pidom}
 \end{equation}
To explore the  possible dependence of the results on  this choice we
also study the  case with  no $\nu_\tau$ production
\begin{equation}
 \left(f_{eS},f_{\mu S},f_{\tau S}\right)= \left(f_{eS}, 1-f_{eS},0\right)
\label{eq:notau}
\end{equation}  
with $\nu_e$ fraction, $f_{eS} \in [0,1]$. 
For the neutrino mixing parameters, we fix the corresponding $
|U_{\alpha i}|$ to the present NuFIT-6.0~\cite{Esteban:2024eli}
best-fit values. We have verified that, within the variations due to
all other uncertainties, the dependence of our results on this choice
is negligible.
\item \emph{Cosmological model}: to compute the Hubble rate $H(z)$, we assume
a $\Lambda$CDM
cosmology with Hubble constant $H_0 = 67.4$\,km\,s$^{-1}$\,Mpc$^{-1}$, and dimensionless
energy density parameters $\Omega_\Lambda = 0.685$ and $\Omega_m =
0.315$~\cite{ParticleDataGroup:2024cfk}.
\item \emph{Normalization constraints}: As mentioned above, both the
  normalization of the source distribution, $\rho_0$; and that of the
  neutrino spectrum at the source, $\Phi_0$; are left to vary in the
  fit, only subject to the constraint in \cref{eq:lim1} and
  \eqref{eq:lim2} for the \emph{Only $\nu$ Attenuation} and
  the \emph{All Particle Attenuation} scenarios, respectively (see
  \cref{eq:norm} below).  These constraints set limits on the
  differential flux times neutrino energy squared without neutrino
  loss. If $\gamma_\mathrm{astro} = 2$, the limits are independent of
  neutrino energy. If, on the contrary, $\gamma_\mathrm{astro}\neq 2$,
  they are meant to apply for all neutrino energies $E$ within the
  range $60\,\mathrm{TeV} \leq E \leq 10^3\,\mathrm{TeV}$. In
  practice, for $\gamma_\mathrm{astro}>2$ this is ensured by choosing
  $E\equiv E_\mathrm{lim}=60$ TeV in \cref{eq:lim1,eq:lim2}; and for
  $\gamma_\mathrm{astro}<2$ this is ensured by choosing $E\equiv
  E_\mathrm{lim}=10^4$ TeV. With our choices we get
\begin{equation}
\Phi_{0} \times \rho_0 \times  \left(\frac{E_\mathrm{lim}}{1\,\rm TeV}\right)^{-\gamma_\mathrm{astro}+2} \times I_0(\gamma_\mathrm{astro})
\leq 5.5 \times 10^{-41} \;(1.1\times 10^{-31}) \;
{\rm GeV^{-1}\, cm^{-3} \, s^{-1}}\;,
\label{eq:norm}
\end{equation}
for the \emph{Only $\nu$ Attenuation} (\emph{All Particle
Attenuation}) scenario. Here, $I_0(\gamma_\mathrm{astro})$ is the
dimensionless integral
\begin{equation}
  I_0(\gamma_\mathrm{astro}) \equiv \int_{z_\mathrm{min}}^\infty
  \frac{(1+z)^{-\gamma_\mathrm{astro}} F(z)}{\sqrt{\Omega_m (1+z)^3 +
      \Omega_\Lambda}}\,\mathrm{d}z \, ,
  \label{eq:I_astro}
\end{equation}
which depends on the choice of source evolution via the function $F(z)$.

To parametrize the dependence of our results on the overall normalization, we
introduce a dimensionless normalization parameter ${\cal N}$
\begin{equation}
  {\cal N}\equiv\frac{ \Phi_{0} \times \rho_0 \times
    \left(\frac{E_\mathrm{lim}}{1\,\rm
      TeV}\right)^{-\gamma_\mathrm{astro}+2} \times
    I_0(\gamma_\mathrm{astro})} {5.6\times 10^{-41} \; {\rm GeV^{-1}\,
      cm^{-3} \, s^{-1}}} \;.
\label{eq:Ndef}
\end{equation}
so that ${\cal N}_{\rm max}=1$ (${\cal N}_{\rm max}=2\times 10^{9}$)
in the \emph{Only $\nu$ Attenuation} (\emph{All Particle Attenuation})
scenario.
\end{itemize}

\section{Flux predictions}
\label{sec:predic}
With all the considerations and choices described in
Sec.~\ref{sec:form}, we obtain that the fluxes arriving at Earth in
the presence of attenuation can be written as
\begin{equation}
  \phi_\alpha= {\cal N} \times
  \left(\frac{E_\mathrm{lim}}{1\,\rm TeV}\right)^{\gamma_\mathrm{astro}-2}
  \sum_{i,\,\beta} |U_{\alpha i}|^2 |U_{\beta i}|^2 f_\beta \left(\frac{E}{1\,\mathrm{TeV}}\right)^{-\gamma_\mathrm{astro}}
  \times \frac{I_\mathrm{att}(\gamma_\mathrm{astro},E, n)}
      {I_0(\gamma_\mathrm{astro})}
\times {6\times 10^{-14} \;
{\rm GeV^{-1}\, cm^{-2} \, s^{-1}} \, {\rm sr}^{-1}} \;, 
\label{eq:attflux}
\end{equation}
where the impact of neutrino loss is to modify $I$ in an energy-dependent
way, $I_0(\gamma_\mathrm{astro}) \rightarrow I_\mathrm{att}(\gamma_\mathrm{astro}, E,n)$, with
\begin{equation}
  I_\mathrm{att}(\gamma_\mathrm{astro}, E,n)
  \equiv \int_{z_\mathrm{min}}^\infty
  \frac{(1+z)^{-\gamma_\mathrm{astro}} F(z)}{\sqrt{\Omega_m (1+z)^3 +
      \Omega_\Lambda}} e^{-\tau(E, z, n)}\, \mathrm{d}z .
  \label{eq:I_att_definition}
\end{equation}

\begin{figure}[h]
    \centering
    \includegraphics[width=0.8\linewidth]{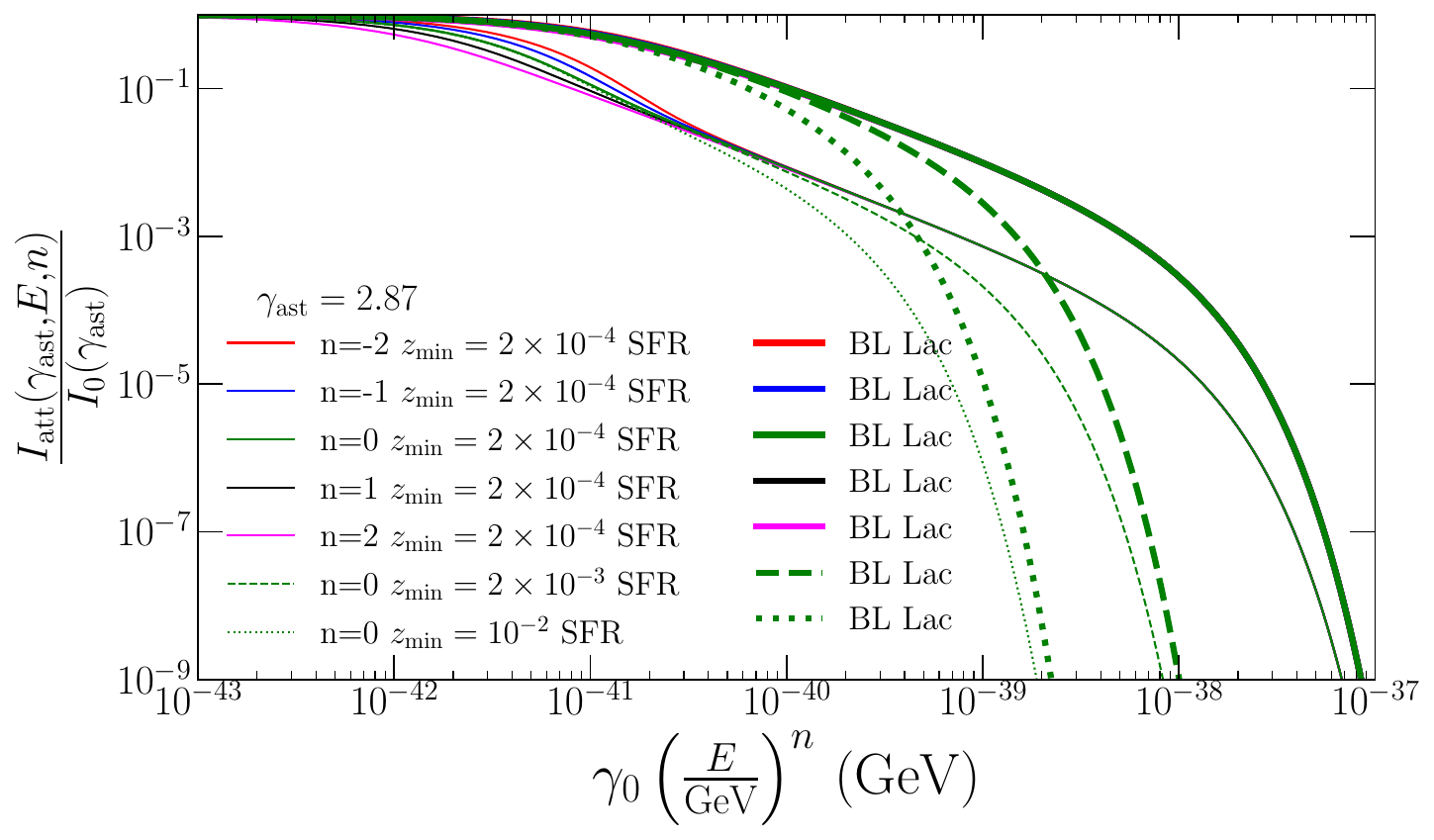}
     \caption{$I_\mathrm{att}/I_0$ as a function of the attenuation exponent
       $\Gamma\equiv\gamma_0 (E/{\rm GeV})^n$ for the different
       parameter choices shown as labeled.  Thinner (thicker) lines
       correspond SFR (BL-Lac) source evolution. The curves for
       different values of $n$ and BL-Lac source evolution totally
       overlap within the resolution of the figure.}
\label{fig:rdec}          
\end{figure}

For reference, we plot in Fig.~\ref{fig:rdec} the ratio
$I_\mathrm{att}/I_0$ as a function of the attenuation exponent
$\Gamma_n\equiv\gamma_0(E/{\rm GeV})^n$ for different parameter
choices as labeled in the figure. In the figure, thin lines correspond
to SFR source evolution while thick lines correspond to BL-Lac source
evolution. As seen in the figure, all other parameters being the same,
the attenuation for SFR source evolution is stronger than that for
BL-Lac evolution. This can be understood by computing the median
redshift $z_\mathrm{med}$ for each source distribution $F(z)$, given
by the median of the integrand in \cref{eq:I_astro}. We find
$z_\mathrm{med}\simeq 0.7$ and $z_\mathrm{med}\simeq 0.09$ for SFR and
BL-Lac source evolution, respectively. Approximating all source
redshifts by $z_\mathrm{med}$, the neutrino flux is attenuated by a
factor $\simeq e^{-\tau(E, z_\mathrm{med})}$, where
\begin{equation}
\begin{split}
\tau(E, z_\mathrm{med}) = \Gamma_n
\int_0^{t_\mathrm{L}(z_\mathrm{med})} \mathrm{d}t_\mathrm{L} (1+z)^n \,
& = \begin{cases}
  [0.28, 0.77] \times \frac{\Gamma_n}{H_0}& \text{for SFR evolution} \\
  [0.077, 0.092] \times \frac{\Gamma_n}{H_0} & \text{for BL-Lac evolution}
\end{cases} \\
& = \begin{cases}
  [0.19, 0.53] \times 10^{42} \times \frac{\Gamma_n}{\mathrm{GeV}}& \text{for SFR evolution} \\
  [0.053, 0.064] \times 10^{42} \times \frac{\Gamma_n}{\mathrm{GeV}} & \text{for BL-Lac evolution}
\end{cases} 
\end{split}\, ,
\label{eq:tau_approx}
\end{equation}
where $t_\mathrm{L}$ is the lookback time defined below
\cref{eq:tau_eff}, and the ranges correspond to varying $n \in [-2,
  2]$. Thus, attenuation is stronger for SFR evolution than for BL-Lac
evolution, as expected since in the former scenario sources are
located at larger redshifts.

The same qualitative argument explains why, as seen in the figure, the
ratio $I_\mathrm{att}/I_0$ mainly depends on the exponent
$\Gamma_n = \gamma_0 (E/\mathrm{GeV})^n$. There is only a mild
dependence on $n$ for SFR source evolution and a very narrow range of
$\Gamma_n$, while for BL-Lac source evolution the dependence on $n$ is
below the resolution of the figure. 

Furthermore, in the figure we fix $\gamma_\mathrm{astro}=2.87$ for
concreteness, but the dependence of the plotted ratio on
$\gamma_\mathrm{astro}$ is very weak. This is, the dependence of the
fluxes on $\gamma_\mathrm{astro}$ is mostly determined by the
unattenuated spectral factor $E^{-\gamma_\mathrm{astro}}$ and the
normalization constant $(E_{\rm lim}/{\rm
  TeV})^{(\gamma_\mathrm{astro}-2)}$. Finally, we show the ratio for
several values of $z_{\rm min}$ as different line styles. As expected,
the larger $z_{\rm min}$ the stronger the attenuation, as sources are
placed at higher redshifts. However, the effect is only relevant for
large values of $\Gamma_n$, where the ratio $I_\mathrm{att}/I_0$ is
well below $10^{-1}$.

\begin{figure}[h]
    \centering
        \includegraphics[width=0.9\linewidth]{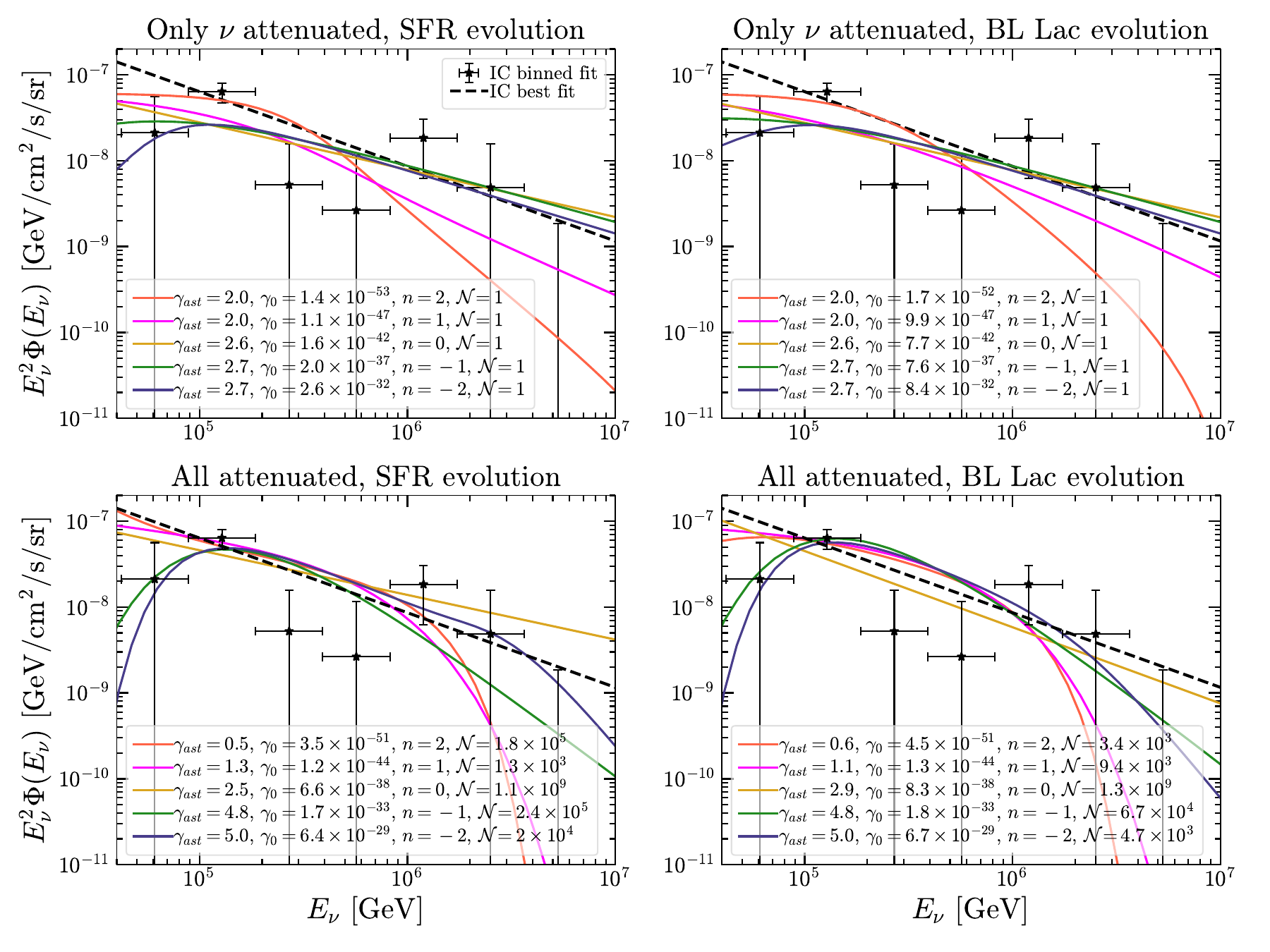}
        \caption{Predicted diffuse neutrino fluxes in the presence of
          neutrino loss for the different scenarios and parameters as
          labeled in the figure. In all cases, we fix the
          normalization to the maximum allowed value, and we assume
          $\pi$-dominance composition, \cref{eq:pidom}.  For
          reference, we show as a dashed black line the best-fit flux
          reported by the IceCube collaboration for their fit with
          $E^{-\gamma_{\rm astro}}$ spectra, as well as their binned
          reconstructed spectra with the corresponding
          uncertainties~\cite{IceCube:2020wum,
            IC75yrHESEPublicDataRelease}.  See text for details.  }
        \label{fig:fluxes}
\end{figure}

Altogether, as illustration of the expected sensitivity, we show in
Fig.~\ref{fig:fluxes} the predicted fluxes for several choices for the
attenuation parameters, source evolution, and universality of the
effect. For reference, we also show the results of the fit performed
by the IceCube collaboration assuming an energy dependence
$E^{-\gamma_{\rm astro}}$ with a constant $\gamma_{\rm astro}$ in the
full energy range ---which results in $\gamma_{\rm astro}=2.87$ (black
dashed line)---, and with $\gamma_{\rm astro}$ fixed to $-2$ but the
normalization allowed to vary independently within each energy bin
(points with error bars).  Let us stress that these IceCube points
with error bars are \emph{not} the data that we are fitting in our
analysis. They correspond to the results obtained by the IceCube
collaboration fitting the same data and with the same systematics
which we are employing (described below), but with a different
assumption on the energy dependence of the fitted spectra. Hence, we
caution that from these error bars alone one cannot conclude whether a
model is allowed or excluded.

\section{Data analysis and results}
\label{sec:results}
In our study we focus on the diffuse neutrino flux, using the IceCube
7.5-year High-Energy Starting Events (HESE)
sample~\cite{IceCube:2013low} to derive quantitative bounds.  The
public 7.5-year IceCube HESE
sample~\cite{IceCube:2020wum,IC75yrHESEPublicDataRelease} that we use
contains 102 detected events with energies between tens of TeV and a
few PeV. A more recent sample~\cite{IceCube:2023sov,
  IC12yrHESEPublicDataRelease}, using 12 years, contains 164 events.
However, because it has no publicly available accompanying sample of
Monte Carlo events that we can use in our statistical analysis, we use
the 7.5-year sample instead.

A HESE event is classified in one of three morphologies---cascades,
tracks, and double cascades---according to its light profile. Each
morphology contains different proportions of the different neutrino
flavours.  To accurately model the capability of IceCube to detect
high-energy astrophysical neutrinos via HESE events, we use the Monte
Carlo (MC) code~\cite{IC75yrHESEPublicDataRelease} that is provided by
the IceCube Collaboration together with the 7.5-year HESE data
release~\cite{IceCube:2020wum}.  In short, we inject in the MC the
neutrino flux at Earth (\cref{eq:attflux}), predicted for a given
value of the model parameters $\vec\omega=({\cal
  N},\gamma_\mathrm{astro},\gamma_0,n,f_\alpha)$.  This generates the
expected number of events as a function of energy, angle, and
morphology carefully accounting for the contribution of the different
sources of backgrounds as well as for the effect of all systematic
uncertainties in the official analysis (here referred to as
$\vec\xi$).

The result is the number of predicted events with morphology $t$
within a given 2D bin $\{i, j\}$, $N^{\rm
  P}_{ij,t}(\vec\omega,\vec\xi)$.  For tracks and cascades, $i$ is an
energy bin, and $j$ an angle bin. For double cascades, $i$ is an
energy bin, and $j$ a bin in separation among cascades. The HESE
sample employed contains 21 bins in energy, evenly spaced in
$\log_{10}(E_\mathrm{dep}/\mathrm{GeV})$, between 60\,TeV and 10\,PeV;
10 bins in directions, evenly spaced in $\cos \theta_z^\mathrm{rec}$,
between -1 and 1; and 20 bins in cascade separation, evenly spaced in
$\log_{10}(\ell/\mathrm{m})$, between 10\,m and 1\,km.

These are compared with observations $N^{\rm O}_{ij,t}$ using a Poissonian
$\chi^2$
\begin{equation}
  \chi^2(\vec\omega)={\rm min}_{\vec\xi}\left[2\sum_{i,j,t}
      \left(N^{\rm P}_{ij,t}(\vec\omega,\vec\xi)
      -N^{\rm O}_{ij,t}+ N^{\rm O}_{ij,t}\ln\frac
        {N^{\rm O}_{ij,t}}{N^{\rm
            P}_{ij,t}(\vec\omega,\vec\xi)}\right)
        +\chi^2(\vec\xi)\right] \, ,
  \label{eq:chisq_def}
\end{equation}
where $\chi^2(\vec\xi)$ is the usual pull term that represents the previous
knowledge of the systematic uncertainly parameters as provided in the
data release~\cite{IC75yrHESEPublicDataRelease}.

We list in Table~\ref{tab:bounds} the 95\% CL bounds on $\gamma_0$ for
different values of the attenuation index $n$ after marginalizing over
${\cal N}$, $\gamma_\mathrm{astro}$ and all systematics; for the
different choices of source evolution and attenuation scenario.
Further details on the statistical procedure and model parameters are given in Appendix \ref{sec:fit-params}.
\begin{table}
\begin{tabular}{|c||c|c|c||c|c|}\hline
  &\multicolumn{5}{c|}{$\gamma_0^{\rm max}$ (GeV) } \\
  \hline
  & \multicolumn{3}{c||} {Only $\nu$  attenuation}
  & \multicolumn{2}{c|} {All particle attenuation } \\\hline
  n&  SN1987A~\cite{Ternes:2025mys}(90\%CL)&
  This work (SFR)  & This work (BL-Lac) &
  This work (SFR) & This work (BL-Lac) \\\hline
   2 & 2.3--3.2 $\times 10^{-34}$ &
  $1.4 \times 10^{-53}$  & $1.7 \times 10^{-52}$& $3.5 \times 10^{-51}$ &
  $4.5 \times 10^{-51}$
  \\
   1 & 1.5--1.8 $\times 10^{-35}$ &
  $1.1 \times 10^{-47}$  & $9.9 \times 10^{-47}$& $1.2 \times 10^{-44}$ &
  $1.3 \times 10^{-44}$  
  \\
  0 & 3.1--5.0 $\times 10^{-37}$ &
  $1.6 \times 10^{-42}$  & $7.7 \times 10^{-42}$& $6.6 \times 10^{-38}$ &
  $8.3 \times 10^{-38}$
  \\
  -1 & 3.2--5.5 $\times 10^{-39}$ &
  $2.0 \times 10^{-37}$  & $7.6 \times 10^{-37}$& $1.7 \times 10^{-33}$ &
  $1.8 \times 10^{-33}$
  \\
  -2  &3.1--6.4 $\times 10^{-41}$ &
  $2.6 \times 10^{-32}$
  & $8.4 \times 10^{-32}$& $6.4 \times 10^{-29}$ &
  $6.7 \times 10^{-29}$  \\\hline
  \end{tabular}
\caption{95\% CL upper bound on the attenuation parameter $\gamma_0$
  from the analysis of the IceCube 7.5-year HESE events for the
  different assumptions about the universality of the attenuation,
  energy dependence, and evolution of neutrino sources (see text for
  details).  Results are show for a reference energy in $E_0=1$ GeV
  (see \cref{eq:Gam}).  We find that, within the precision of the
  bounds quoted, they hold for both a $\pi$-decay dominance flavour
  composition,~\cref{eq:pidom}; as well as for the most general
  no-$\nu_\tau$ scenario,~\cref{eq:notau}.  For comparison, we show
  the 90\% CL bounds from the analysis of SN1987A data in
  Ref.~\cite{Ternes:2025mys}, which only apply to the \emph{Only $\nu$
  Attenuation} scenario.}
\label{tab:bounds}
\end{table}

The characteristic values of the bounds and their dependence on $n$ can
be understood in terms of the semi-quantitative arguments provided in
the previous section.  First, from Fig.~\ref{fig:fluxes}, we notice
that the IceCube binned best-fit fluxes practically reach the WB bound
at $E\sim 100$ TeV, while they are about one order of magnitude below
the bound at $E\sim 1000$ TeV.

In the \emph{Only $\nu$ attenuation} scenario, this requires the ratio
plotted in Fig.~\ref{fig:rdec} to be $\lesssim 1$ for $E \sim
10^5\,\mathrm{GeV}$, and $\lesssim 0.1$ for $E \sim
10^6\,\mathrm{GeV}$.  If $n \leq 0$, the first constraint is the
strongest, and from Fig.~\ref{fig:rdec} it implies $\gamma_0 \,
(10^5\,\mathrm{GeV}/\mathrm{GeV})^n = \gamma_0 \, 10^{5n} \lesssim
\mathrm{few} \times 10^{-42}\,\mathrm{GeV}$ for SFR source evolution,
and $\gamma_0 \, 10^{5n} \lesssim \mathrm{few} \times
10^{-41}\,\mathrm{GeV}$ for BL-Lac source evolution. If, in turn, $n >
0$, the second constraint is the strongest. For SFR source evolution,
it implies $\gamma_0 \, (10^6\,\mathrm{GeV}/\mathrm{GeV})^n = \gamma_0
10^{6n} \lesssim \mathrm{few} \times 10^{-41}$---or, equivalently,
$\gamma_0 \, 10^{5n} \lesssim \mathrm{few} \times
10^{-41-n}\,\mathrm{GeV}$. For BL-Lac source evolution, it implies
$\gamma_0 \, 10^{5n} \lesssim \mathrm{few} \times
10^{-40-n}\,\mathrm{GeV}$. (Similar conclusions follow from
\cref{eq:tau_approx}.)

\begin{figure}[h]
\centering
\includegraphics[width=0.6\linewidth]{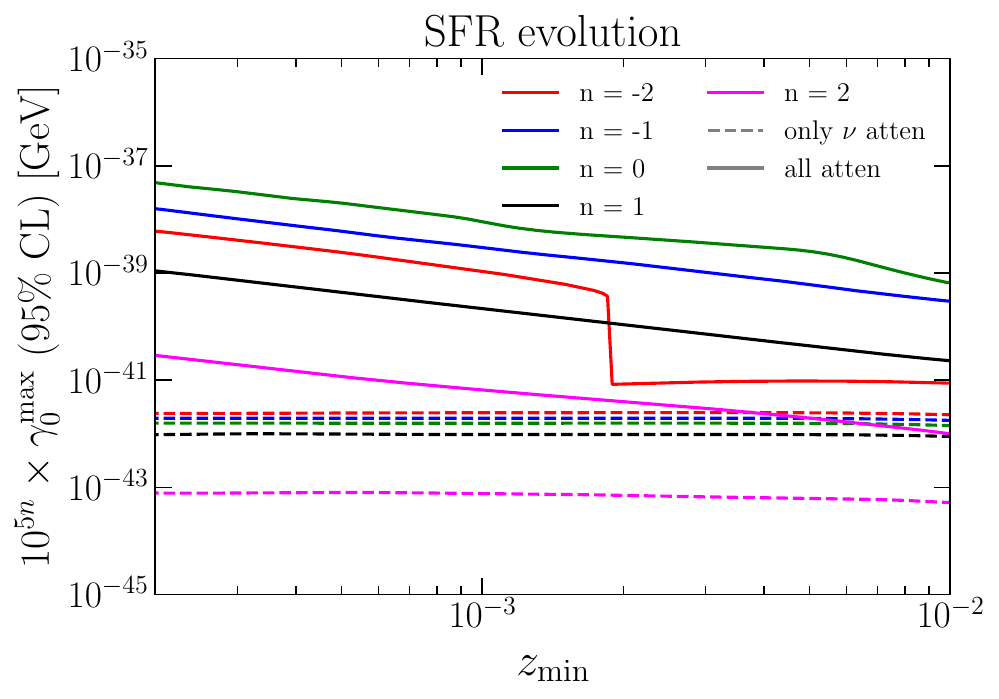}
\caption{Dependence of the 95\% CL $\gamma_0$  bound on the minimum redshift
for the different attenuation indexes $n$ and scenarios (see text for details).}
\label{fig:zmin}
\end{figure}

From Fig.~\ref{fig:rdec}, we expect these bounds to be quite
independent of the minimum redshift considered. This is explicitly
displayed in the dashed lines in Fig.~\ref{fig:zmin}, where we show
the 95\%CL bounds on $\gamma_0$ in this scenario as a function of
$z_{\rm min}$ for different attenuation indices $n$.  For the sake of
concreteness, we show the result for SFR source evolution but the
behavior for BL-Lac is very similar.  We furthermore find that, under
our working assumption that neutrino loss is flavour-universal,
\cref{eq:Gam}, the bounds derived are independent of the flavour
composition scenarios considered within the precision quoted.

For comparison, we list in Table~\ref{tab:bounds} the 90\%CL bounds
reported in Ref.~\cite{Ternes:2025mys} from the analysis of SN1987A
neutrino data under different model assumptions for the supernova
neutrino fluxes. As the modeling of the supernova relies on the
observed light, strictly, these bounds only apply in an scenario where
only the neutrinos are attenuated. Given the characteristic MeV
energies of supernova neutrinos, those bounds are only stronger than
IceCube constraints for attenuation indexes $n<0$, for which the
effect is larger at lower energies.

Moving now to the \emph{All Particle Attenuation} scenario, in this
case ${\cal N}$ can be $2\times 10^{9}$ times larger than for the WB
bound. Thus the suppression factor is only bounded to be $\lesssim 5
\times 10^{-9}$--$5 \times 10^{-10}$. As seen from
Fig.~\ref{fig:rdec}, for our choice $z_{\rm min}=2 \times 10^{-4}$,
this requires $\Gamma_n\lesssim \mathrm{few} \times 10^{-38}$ GeV for
either source evolution. Following the same reasoning as above, this
implies $\gamma_0/{\rm GeV}\times 10^{5n}\lesssim {\rm few}\times
10^{-38}$ for $n \leq 0$ and $\gamma_0/{\rm GeV}\times 10^{5n}\lesssim
{\rm few}\times 10^{-38-n}$ for $n > 0$.

We find a very mild dependence of the bounds on the assumed source
evolution, as expected from Fig.~\ref{fig:fluxes}. The dependence of
the bound with $z_{\rm min}$ is shown in the full lines in
Fig.~\ref{fig:zmin}.  As expected in this case, the bounds become
stronger as $z_{\rm min}$ increases, with a ``jump'' for $n=-2$ that
we discuss below.

\begin{figure}[h]
    \centering
        \includegraphics[width=0.9\linewidth]{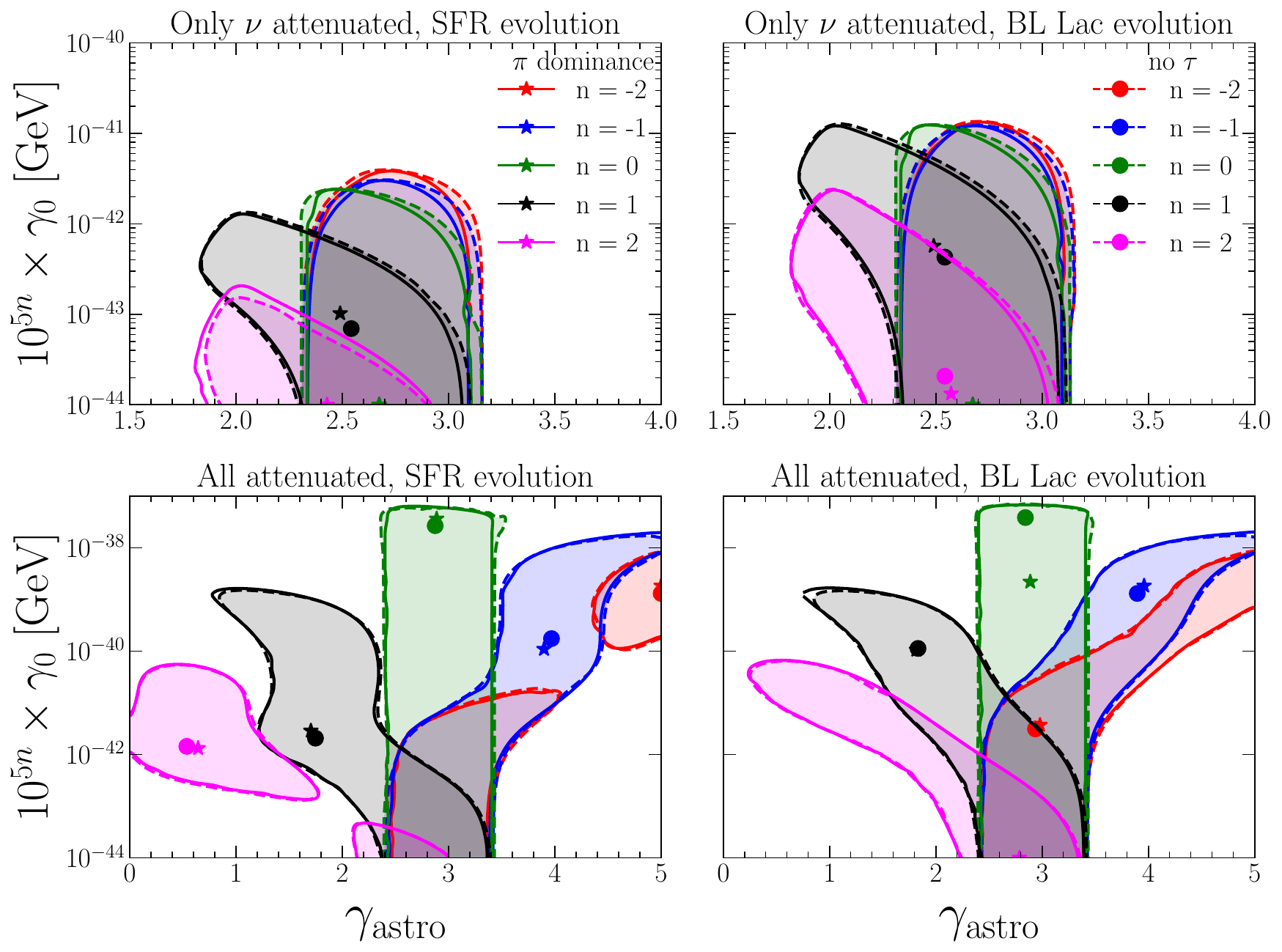}
        \caption{95\% CL allowed regions (2dof) in the ($\gamma_{0},
          \gamma_\mathrm{astro}$) plane after marginalizing over
          ${\cal N}$ and all systematic uncertainties, for different
          choices of the attenuation index $n$, the attenuation
          scenario, and source evolution, as labeled in the
          figure. The solid (dashed) contours are the results assuming
          a flavour composition of $\pi$-domination,~\cref{eq:pidom}
          (no-$\nu_\tau$,~\cref{eq:notau}). In all cases, $z_{\rm
            min}=2 \times 10^{-4}$.}
        \label{fig:gammas}
\end{figure}

Our results also show an interesting interplay between the allowed
attenuation factor and the best-fit spectral index. This is
illustrated in Fig.~\ref{fig:gammas}, where we show the results of the
analysis in the form of the 95\% CL allowed region for the attenuation
parameter $\gamma_0$ and the flux spectral index
$\gamma_\mathrm{astro}$, for different values of the attenuation index
$n$. We marginalize over ${\cal N}$ and all systematic uncertainties,
and we show the results for the different choices of source evolution,
universality of the attenuation, and flavour composition (see caption
for details).

From the figure, we see a clear correlation between the allowed ranges
of $\gamma_0$ and $\gamma_\mathrm{astro}$ when attenuation is large,
in particular for the \emph{All Particle Attenuation} scenario. This
can be understood from Fig.~\ref{fig:rdec}: for large values of
$\Gamma_n = \gamma_0 (E/\mathrm{GeV})^n$, but before the
$z_\mathrm{min}$ cutoff, $I_\mathrm{att}/I_0$ is approximately
proportional to $1/\Gamma_n \propto \frac{1}{\gamma_0}
E^{-n}$. Physically, exponential attenuation strongly suppresses the
neutrino flux of sources located at a light-travel time $\gg
1/\Gamma_n$, making the flux at Earth proportional to $\sim
1/{\Gamma_n}$ (in the same way that the Olbers paradox dictates the
brightness of the sky to be proportional to the radius of the
observable Universe) --- this approximate proportionality can also be
explicitly obtained from \cref{eq:I_att_definition}. In this range of
parameters, the attenuated flux has therefore a spectral dependence
\begin{equation}
  \phi_\mathrm{att}(E)\sim  \frac{1}{\gamma_0} E^{(-\gamma_\mathrm{astro}-n)}\, .
\label{eq:qdeg}  
\end{equation}
Since the HESE data can be well fit with an non-attenuated power-law
flux with best fit
$\gamma_\mathrm{astro}=2.87$~\cite{IceCube:2020wum}, including
attenuation in this regime one can obtain a fit of the same quality if
$\gamma_\mathrm{astro}=2.87-n$.  This makes the allowed regions extend
to lower values of $\gamma_\mathrm{astro}$ for $n>0$, and to higher
values of $\gamma_\mathrm{astro}$ for $n<0$; as can be seen in the
lower panels in Fig.~\ref{fig:gammas}. In some cases, the best fit
around $\gamma_\mathrm{astro}=2.87-n$ is slightly better than the
``standard'' one (corresponding to $\gamma_\mathrm{astro}=2.87$,
$\gamma_0\rightarrow 0$), which leads to the disconnected 95\% CL
regions observed in the lower left panel. As can be seen in
Fig.~\ref{fig:rdec}, this quasi-degeneracy is cut off by the effect of
$z_{\rm min}$.  If one increases $z_{\rm min}$, the range of
parameters for which the second solution is possible disappears and
only the standard minimum remains. For $n=-2$, this leads to the
``jump'' in the $z_{\rm min}$ dependence of the bound displayed in the
red curve in Fig.~\ref{fig:zmin}.

\section{Summary and conclusions}
\label{sec:conclu}
In this work we have explored the sensitivity of the observed diffuse
flux of high-energy neutrinos to new physics resulting into an
exponential attenuation of the flux. Working in a model-independent
form, we have employed a generic parametrization of the effect,
\cref{eq:Gam}, allowing for an arbitrary attenuation length
parametrized by the attenuation factor $\gamma_0$ and different energy
dependences parametrized by the attenuation index $n$.  Qualitatively,
our results rely on the observation that conservative
energy-conservation arguments allow to severely constrain such
neutrino losses. To that end, we have considered two possible
energy-conservation limits. On the one hand, \emph{Only $\nu$
Attenuation}, where the attenuation only affects neutrinos while
leaving all other messengers unaffected. This allows to employ the
so-called Waxman-Bacall bound, \cref{eq:lim1}, to set limits on
neutrino attenuation. On the other hand, the more extreme case of
\emph{All Particle Attenuation}, with attenuation affecting all cosmic
messengers. This can still be constrained under the extremely
conservative assumption in \cref{eq:lim2}.

The study requires integrating over the redshift of the many
unresolved sources contributing to the flux. We have considered two
characteristic source evolutions, either following the star formation
rate that places most sources around $z = 1$, \cref{eq:sfr}; or
following the distribution of BL-Lac objects that places more sources
closer to Earth, \cref{eq:bl-lac}. Regarding flavour composition, we
have considered the standard $\pi$ dominated
composition,~\cref{eq:pidom}; and the most general no-$\nu_\tau$
case,~\cref{eq:notau}. We have illustrated the predicted fluxes for
several values of all the parameters and choices considered in the
study in Fig.~\ref{fig:fluxes}.

In all these scenarios, we have performed a careful reanalysis of the
7.5-year HESE samples of IceCube, closely following the analysis
performed by the collaboration for the standard unattenuated
scenario. We have allowed for an arbitrary spectral index of the
unattenuated flux, $\gamma_\mathrm{astro}$; and we have included all
uncertainties and systematics considered by the collaboration. With
that, we have derived the 95\% CL bounds compiled in
Table~\ref{tab:bounds}. In brief, we find that for energy-independent
or energy-growing attenuation (i.e., $n\geq 0$), the bounds derived
from this analysis are stronger than the strongest bound in the
literature derived from SN1987A neutrinos. Furthermore, the bounds are
only degraded by a factor ${\cal O}(1000)$ in the extreme \emph{All
Particle Attenuation} scenario, where the unattenuated neutrino energy
density would exceed observations by more than 9 orders of magnitude. The
dependence of the results on the source evolution we considered is
very mild, and the dependence on the flavour composition marginal.

Variations of the bounds from other astrophysical uncertainties,
such as source spectral breaks or mixed source populations, are
expected to be small. Regarding spectral breaks, as shown in
Ref.~\cite{IceCube:2020wum}, these models provide comparable
predictions to a single power law, so the impact of our
order-of-magnitude bounds on exponential attenuation should be
comparable (see also Ref.~\cite{Valera:2024buc}). Regarding mixed
source populations, as \cref{tab:bounds} shows, we obtain
comparable limits if the sources are located at high redshifts
(SFR) or closer to Earth (BL-Lac).

In addition, our bounds have been derived under the assumption of
neutrino loss being flavour independent of neutrino loss. As mentioned
in the Introduction, our bounds can be still conservatively applied to
flavor-dependent scenarios as long as they are imposed on the lowest
$\gamma_0$. Therefore, if some neutrino states have $\gamma_{0,i}=0$,
the total HESE flux will only be reduced by factors of ${\cal O}(1)$,
and the bounds we derive assumming the total flux attenuation do not
apply. In this case, flavour information becomes paramount to impose
meaningful bounds (see for example Ref.~\cite{Valera:2024buc}).  

Our results also show an interesting interplay between the allowed
attenuation factor and the best-fit spectral index (see
Fig.~\ref{fig:gammas}). This is a consequence of the similar energy
dependence induced by both when large enough attenuation is allowed,
\cref{eq:qdeg}, ---as it is the case in the \emph{All Particle
Attenuation} scenario--- that leads to a quasi-degeneracy.  This
behavior is cut-off by the fact that there is a minimum redshift for
the sources of the neutrinos, which we conservatively set to $2\times
10^{-4}$ (see Fig.~\ref{fig:zmin}). Consequently, meaningful
constraints can still be imposed on both parameters even if the
allowed range of spectral indices is significantly enlarged.  Let us,
however, stress that the \emph{All Particle Attenuation} scenario is
considered as an extreme case to show that even in this scenario the
diffuse neutrino fluxes can provide a bound. Severe constraints should
also follow in this scenario from the observations of the other cosmic
messengers. Nevertheless, as we have argued, even the mere observation of
astrophysical neutrinos allows to set quantitative meaningful limits.

\section{Acknowledgements}
\label{sec:acknow}
This project is funded by USA-NSF grant PHY-25104424. It has also
received support from the European Union's Horizon Europe research and
innovation programme under the Marie Sk\l odowska-Curie Staff Exchange
grant agreement No 101086085 -- ASYMMETRY''. It also receives support
from grants PID2019-105614GB-C21, PID2022-136510NB-C33,
PID2024-156016NB-I00, and ``Unit of Excellence Maria de Maeztu'' award
to the ICC-UB CEX2024-001451-M funded by
MICIU/AEI/10.13039/501100011033 and, as appropriate, by
``ERDF/EU''. It has also been supported by the Basque Government
IT1628-22 grant and the UPV/EHU EHU-N25/11 grant. It has also been
supported by the {\it Dirección de Fomento de la Investigación} at
PUCP, through grant No. CAP PI 1144. Part of this work used the
Solaris cluster, acquired through the Basque Government IT1628-22
grant. GZ gratefully acknowledges funding from the Vicerrectorado de
Investigación at Pontificia Universidad Católica del Perú via the
Estancias Posdoctorales en la PUCP 2023 program. AMG and GZ want to express
their gratitude to the Institut de Ci\`encies del Cosmos Universitat de
Barcelona for their warm hospitality during the development of this
work.


\appendix

\section{Statistical analysis}
\label{sec:fit-params}

Our statistical treatment is a modification of the
one in the public 7.5-year IceCube HESE
sample~\cite{IceCube:2020wum,IC75yrHESEPublicDataRelease}, where the
original power-law flux is replaced by the flux in
Eq.~\eqref{eq:attflux}.  In more detail, we have carried out a
frequentist analysis with model parameters divided into parameters
of interest $\vec{\omega}$ and nuisance parameters $\vec{\xi}$.  Our
parameters of interest are the normalization parameter
$\mathcal{N}$, defined in \cref{eq:Ndef}; the spectral index of the
power-law flux $\gamma_{\rm astro}$; the neutrino loss parameter
$\gamma_{0}$; the energy-dependence index of neutrino loss $n$; and
the $\nu_e$ fraction at sources $f_{e S}$, defined in
\cref{eq:notau}.  The range of the normalization parameter
$\mathcal{N}$ depends on the normalization constraints (\emph{Only
$\nu$ Attenuation} or \emph{All Particle Attenuation}), the
energy-dependence index of neutrino loss is fixed to $n = \{ -2, -1,
0, 1, 2 \}$, and the $\nu_e$ fraction is fixed to $f_{e S} = 1/3$ in
the scenario of $\pi$-dominance, whereas it is a free parameter in
the no $\tau$ scenario.  The nuisance parameters $\vec{\xi}$ are the
same as in the original analysis of the 7.5-year IceCube HESE
sample~\cite{IceCube:2020wum,IC75yrHESEPublicDataRelease} (we refer
the reader to that paper for more details). In \cref{tab:params}, we
list all model parameters with their respective prior constraints
and scan ranges.

\begin{table}[]
    \centering
    \begin{tabular}{|l c c r|}
    \hline
    Parameter & Constraint & Range & Description \\
    \hline
    \textbf{Astrophysical neutrino flux}: & & & \\
    $\mathcal{N}$          & - & $[10^{-15}, 1]$ ($[10^{-15}, 2 \times 10^9]$)   & Normalization parameter \\
    $\gamma_{\rm astro}$        & - & $[0, 5]$    & Spectral index \\
    $\gamma_{0}$           & - & $[10^{-55}, 10^{-25}]$       &
    Neutrino loss parameter \\
    $f_{e S}$                   & - & $[0, 1]$       & Fraction of $\nu_e$ at sources \\
    \hline
    \textbf{Atmospheric neutrino flux}: & & & \\
    $\Phi_{\rm conv}$                     & $1.0 \pm 0.4$ & $[0, \infty)$ & Conventional normalization scale \\
    $\Phi_{\rm prompt}$                   & -             & $[0, \infty)$ & Prompt normalization scale \\
    $R_{K/\pi}$                           & $1.0 \pm 0.1$ & $[0, \infty)$ & Kaon-Pion ratio correction \\
    $2 \nu /(\nu + \bar{\nu})_{\rm atmo}$ & $1.0 \pm 0.1$ & $[0,2]$       & Neutrino-antineutrino ratio correction \\
    \hline
    \textbf{Cosmic-ray flux}: & & & \\
    $\Delta \gamma_{\rm CR}$       & $0.0 \pm 0.05$ & $(-\infty, \infty)$ & Cosmic-ray spectral index modification \\
    $\Phi_{\mu}$                   & $1.0 \pm 0.5$  & $[0, \infty)$       & Muon normalization scale \\
    \hline
    \textbf{Detector}: & & & \\
    $\epsilon_{\rm DOM}$       & $0.99 \pm 0.1$ & $[0.80, 1.25]$  & DOM efficiency \\ 
    $\epsilon_{\rm head-on}$   & $0.0 \pm 0.5$  & $[-3.82, 2.18]$ & Head-on efficiency \\ 
    $a_{\rm s}$                & $1.0 \pm 0.2$  & $[0.0, 2.0]$    & Ice anisotropy scale \\
    \hline
    \end{tabular}
    \caption{\label{tab:params} Model parameters used in our analysis,
      with their respective prior constraints and scan ranges. For
      parameters with prior Gaussian constraints, we provide the means
      and standard deviations; whereas parameters without prior
      constraints are denoted by $-$. For the normalization parameter
      $\mathcal{N}$ we show the scan ranges for the \emph{Only $\nu$
      Attenuation} (\emph{All Particle Attenuation}) scenario. For the
      \emph{$\pi$-dominance} scenario, the parameter $f_{e S}$ is
      fixed to $1/3$ whereas it is a free parameter for the \emph{no
      $\tau$} scenario.}
\end{table}

We find best-fit parameters by minimizing a
Poissonian $\chi^2$ (\cref{eq:chisq_def} in the main text). To this
purpose, we use the L-BFGS-B algorithm~\cite{doi:10.1137/0916069}.
We present our results assuming Wilks' theorem \cite{Wilks:1938dza}
with the corresponding degrees of freedom.

\clearpage

\bibliography{references}

\end{document}